\documentclass[fleqn,10pt]{wlscirep}
\usepackage[utf8]{inputenc}
\usepackage[T1]{fontenc}
\usepackage{array}
\usepackage{booktabs}
\usepackage{tabu}
\usepackage{dcolumn}
\usepackage{subfigure}
\usepackage{graphicx}
\usepackage{dcolumn}
\usepackage{bm}
\usepackage{slashed}
\usepackage{color}
\usepackage[normalem]{ulem}
\usepackage{babel}
\usepackage{amsmath}
\usepackage{amssymb}
\usepackage{graphicx}
\usepackage{esint}
\usepackage{cleveref}
\crefname{equation}{Rel.}{Rels.} 
\crefrangelabelformat{equation}{(#3#1#4--#5#2#6)}
\usepackage{mathtools}
\makeatletter
\@ifundefined{textcolor}{}
{%
    \definecolor{BLACK}{gray}{0}
    \definecolor{WHITE}{gray}{1}
    \definecolor{RED}{rgb}{1,0,0}
    \definecolor{GREEN}{rgb}{0,1,0}
    \definecolor{BLUE}{rgb}{0,0,1}
    \definecolor{CYAN}{cmyk}{1,0,0,0}
    \definecolor{MAGENTA}{cmyk}{0,1,0,0}
    \definecolor{YELLOW}{cmyk}{0,0,1,0}
}
\usepackage{color}\usepackage{euscript}\usepackage{epsfig}
\usepackage{amsfonts}\usepackage{fancyhdr}

\title{Metric Field as Emergence of Hilbert Space}

\author[*]{Maysam Yousefian}
\author[**]{Mehrdad Farhoudi}
\affil[ ]{Department of Physics,
    Shahid Beheshti University, 1983969411, Tehran, Iran}
\affil[*]{M\_Yousefian@sbu.ac.ir}
 \affil[**]{m-farhoudi@sbu.ac.ir}

\begin{abstract}
First, we explain some ambiguities of spacetime and metric field
as fundamental concepts. Then, from the Unruh effect point of view
and using the Gelfand-Naimark-Segal construction, we construct an
operator as a quanta of acceleration that we call quantum
acceleration operator (QAO). Thereupon, we investigate the
relation between the vacuum of two different frames in the
Minkowski space. Also, we show that the vacuum of each accelerated
frame in the Minkowski space can be obtained by applying such a
QAO to the Minkowski vacuum. Furthermore, utilizing these QAOs, we
augment the Hilbert space and then extract the metric field of a
general frame of the Minkowski spacetime. In this approach, these
concepts emerge from the Hilbert space through the constructed
QAOs. Accordingly, such an augmented Hilbert space includes
quantum field theory in a general frame and can be considered as a
fundamental concept instead of the classical metric field and the
standard Hilbert space.
\end{abstract}
\begin{document}

\flushbottom
\maketitle
\thispagestyle{empty}

\section*{Introduction}

The Einstein theory of gravitation has been one of the most
elegant and interesting theories of physics with a robust
mathematical framework, see, e.g., Refs.~\cite{Janssen2005,FarYou}
and references therein. In addition, being independent from
observer is one of its most important aspects. It has
significantly altered our perception of the natural world, and
even its formalism and concepts have largely formulated the
standard understanding of physical laws. Indeed, his theory has
been universally accepted as the best theory to describe
gravitation at the classical level and provides acceptable
predictions in explaining some cosmic
phenomena\rlap.\cite{Will2006} Nevertheless, despite its dominant
successes, due to a number of seeming insufficiencies and
shortcomings, it would be perceived as a step towards a much more
complete and comprehensive structure, see, e.g.,
Ref.~\cite{Ishak2019}. In this regard, many alternative attempts
have been made for various
modifications/generalizations/amendments of general relativity
(GR), see, e.g.,
Refs.~\cite{Farhoudi2006,Farhoudi2009,Sotiriou2010,
CapoLaur,Clifton, Joyce2015,zare3,Quiros2018,Shiravand2022} and
references therein.

In GR, the metric tensor usually plays the role of gravitational
potential field. More importantly, it gives rise to an isomorphism
between tangent and cotangent spaces, which simply means that it
lowers and raises tensorial indices. Moreover, the metric field as
a property of spacetime describes the geometrical and causal
structure of spacetime and is used to define concepts such as
time, distance, volume, curvature and angle, and to separate
future tenses from past tenses. In addition, in a manifold endowed
with metric, it is possible to compare `lengths' in different
(metric) geodesic curves and identify such a parameter, e.g., with
the proper time. In this regard, to provide a correct physical
description of gravitation, there must first be a correct physical
description of spacetime. Meanwhile, it should be noted that the
current description of GR for spacetime is a mathematical model
that is~not physically observable. The reason is that according to
this theory and quantum mechanics, if we want to observe the
structure of spacetime more precisely, particles with more energy
will be needed, which itself causes more curvature of spacetime.
Such an iterative process of increasing curvature continues until
a black hole is formed. However, when a black hole is formed, the
exact structure of spacetime can no~longer be observed.
Accordingly, it seems necessary to redefine spacetime and/or
metric tensor in a more physically acceptable way.

On the other hand, it is known that the Hilbert space of states
supervenes spacetime. It is a philosophical expression in the
sense that a fact or a property entails or results from the
existence or creation of another. In general, the latter is a
lower-level phenomenon and the former is a higher-level one.
Indeed, the mathematical structure and physics of quantum
mechanics is silent about where the Hilbert space exists, but
nowhere. This is a kind of ontological puzzle that overshadows
speculations not~only on the higher spatial dimensions in string
theory, but even on the number of ordinary dimensions itself.
Furthermore, it is quite clear that everything we observe and
measure in nature is done by elementary particles, which are
themselves explained through the theory of quantum mechanics and
quantum field theory (QFT). That is, the fundamental ontology of
the universe is fully represented by a vector in an abstract
Hilbert space that evolves in time according to the unitary
Schr\"odinger dynamics. Accordingly, everything else, from fields
and particles to space itself, can properly emerge from a set of
Hilbert space ingredients. This approach is called mad-dog
Everettianism\rlap.\cite{Mad-dog}

Another phenomenon that challenges the concept of spacetime as a
fundamental quantity is the phenomenon of entanglement, which has
recently been proven to exist and has also been awarded the Nobel
prize\rlap.\cite{Aspect2022} In this phenomenon, the entangled
particles behave as a single whole and no spacetime distance
between them is considered to establish a connection with each
other. This is such that the effect of any change in one of these
particles can be instantly seen in the other particles associated
with it\rlap.\cite{pas-2023}

Meanwhile, it has recently been suggested in some
researches~\cite{Saravani-2016,Perche-2022} that the metric tensor
as well as the features of spacetime can be obtained from a
two-point correlation function as the inner product of two Hilbert
space vectors. In Ref.~\cite{Perche-2022}, the vacuum state of a
curved spacetime is defined by an annihilation operator of the
wave function, which is the solution of the wave equation based on
the classical metric field. Indeed, to obtain such a wave
function, first the classical metric is employed, and then the
classical metric field is the object through which the quantum
vacuum state of the curved space is derived [From a
mathematical point of view, it can be argued that such a process
represents a tautological loop. However, the method introduced in
Ref.~\cite{Perche-2022} is used to define local quantum clocks and
rulers.].
 Accordingly, such
attempts to explain nature are made using two incompatible bases,
one of which is the classical metric field and the other is the
Hilbert space in quantum mechanics. These two bases are
incompatible because no acceptable standard quantum expression for
the classical metric field has yet been provided.

Nevertheless, in this case, with the aid of the mechanism of the
Unruh effect\rlap,\cite{Unruh-1976} if the Hilbert space can be
augmented to encompass not~only the vacuum of the Minkowski frame
but also the vacuum of any frame in the Minkowski space, the
metric field and consequently the features of spacetime can be
extracted from such an augmented Hilbert space through the
mechanism described in Refs.~\cite{Saravani-2016,Perche-2022}. In
this manner, nature can be described only on the basis of quantum
mechanics, and a quantum description of acceleration can also be
provided. The quantum description of acceleration is significant
because, based on observations related to the local equivalence of
gravitational acceleration and the inertial acceleration, it can
lead us toward a deeper understanding of quantum gravity.

The work is organized as follows. First, in the next section, from
the Unruh effect point of view and using the Gelfand-Naimark-Segal
(GNS) construction\rlap,\cite{Kadison-1983} we review and
investigate the relation between the vacuum of different frames.
Thereupon, in Sec.~III, we search the properties and relation
between the Minkowski vacuum and the vacuum of a stationary rigid
reference frame (SRRF) in the Minkowski space. In this regard, we
recall that, in Ref.~\cite{YousefFarh5}, we presented a somewhat
incomplete draft of the relation between the Rindler and the
Minkowski frames. Then here, in Sec.~IV, we scrutinize the
connection between the Minkowski vacuum and the vacuum of a
general coordinate frame in the Minkowski space. Also, we show how
to augment the standard Hilbert space. In this way, some kind of
quantum expression for acceleration can be researched. In Sec.~V,
we investigate the possibility of extracting the metric field
using tools that will be provided in such an augmented Hilbert
space. In this manner, we try to find a quantum expression as an
alternative to the classical concept of the metric field. Also, in
this section, we try to replace some traditional fundamental
concepts of physics with another fundamental concept. Finally, we
furnish the conclusion in the last section.

\section*{A Vacuum From A Vacuum}
In the arena of QFT in curved spacetimes or in non-inertial
frames, in addition to the review paper by Eguchi et
al.~\cite{Eguchi-1980}, there are also standard textbooks, e.g.,
Refs.\rlap.\cite{Birrell1982,Fulling1989,Wald-1994,
Mukhanov-2007,Parker-2009} However, in this work, we consider a
Hermitian scalar field theory that satisfies the Klein-Gordon
equation according to Ref.~\cite{Crispino-2008}, while following
the Unruh effect approach.

First, we want to obtain QFT in one frame of coordinates, say
$(2)$, in terms of the Fock space in another frame of coordinates,
say $(1)$, under their change of coordinates
$x_{(2)}^\mu=x_{(2)}^\mu(x_{(1)})$. In this case, each
$x_{(2)}^\mu$ is a single-valued continuous function of all
coordinates $(1)$ (at least for certain ranges of their arguments)
and the lowercase Greek indices run from zero to three. For this
purpose, a general solution scalar field, say $ \hat{\Phi} $ with
mass $m$, as the superposition of the positive-frequency mode
bases solutions $f_{i_s}^{(s)} $ to the corresponding Klein-Gordon
equation for $ s=1$ and/or $2$, can be written as
\begin{equation}\label{GeneralField}
        \hat{\Phi}(x)=\sum_i\left[\hat{a}_{i_s}^{(s)}f_{i_s}^{(s)}+\hat{a}_{i_s}^{(s)\dagger}f_{i_s}^{(s)\dagger}
        \right].
\end{equation}
In \cref{GeneralField}, each $\hat{a}_{i_s}^{(s)} $ is an
annihilation operator in the corresponding frame with the standard
commutation relations.

Considering the commutation relations and noting that every
complex conjugate of such a field (as an
involution\cite{Bratteli-1987,Kadison-1983}) is actually a
Klein-Gordon field, we can conclude that the Klein-Gordon algebra
qualifies as a $^*$-algebra. Furthermore, since the Klein-Gordon
field is a vector space, we define a map as the norm function for
this field as $\parallel\hat{\Phi}\parallel_\xi\:\:
:=\sqrt{<\xi|\hat{\Phi}\hat{\Phi}^\dagger|\xi>}$, where $\xi$ is a
normal vector of the Hilbert space. Such a norm function defines a
metric topology in the Klein-Gordon field and this field is
complete with respect to this topology. Based on the
characteristics of the norm function, we can conclude that the
Klein-Gordon field also qualifies as a
C$^*$-algebra\rlap.\cite{Bratteli-1987} Moreover, utilizing the
GNS construction, we can express the expectation value $\Phi_\xi$
of the field $\hat{\Phi}$, while using the $^*$-representation
$\pi\: :\: \hat{\Phi}\rightarrow\hat{\Phi}$, as
$\Phi_\xi:=<\xi|\hat{\Phi}|\xi>$.

Now to obtain QFT in coordinates $(2)$ in terms of the Fock space
in coordinates $(1)$, based on the proposition $4.5.3$ of
Ref.~\cite{Kadison-1983}, we need to gain an isomorphism $U$ from
the Hilbert space $\mathcal{H}_{(1)}$ onto the Hilbert space
$\mathcal{H}_{(2)}$. For this purpose, the solutions and operators
of these two frames are related to each other via the Bogoliubov
transformation~\cite{Crispino-2008,Bogoliubov-1958} as
\begin{equation}\label{FrameiToFrameI}
        f_{i_{2}}^{(2)}=\sum_{i_1}\left[\alpha_{i_{2}i_1}f_{i_1}^{(1)}+\beta_{i_{2}i_1}f_{i_1}^{(1)*}\right]
\qquad\ \text{and}\qquad\
  \hat{a}_{i_{2}}^{(2)}=\sum_{i_1}\left[\alpha_{i_{2}i_1}^*\hat{a}_{i_1}^{(1)}-\beta_{i_{2}i_1}^*\hat{a}_{i_1}^{(1)\dagger}\right]
\end{equation}
and vice versa, where $\alpha_{i_{2}i_1} $ and $\beta_{i_{2}i_1} $
and their conjugates are the Bogoliubov
coefficients\rlap.\cite{Bogoliubov-1958} These coefficients can be
used to obtain the average total number of particles in frame
$(2)$ with energy $i_2$, say $ N_{i_2}^{(2)} $, from the vacuum
state of frame $(1)$. In this way, using \cref{FrameiToFrameI}, we
obtain
\begin{equation}\label{TotalNumber}
N_{i_2}^{(2)}=<0^{(1)}|\hat{a}_{i_2}^{(2)\dagger}\hat{a}_{i_2}^{(2)}|0^{(1)}>=\text{tr}\left(\beta_{i_2i_1}\beta_{i_2j_1}^*\right).
\end{equation}
Also, by inserting \cref{FrameiToFrameI} into the standard
commutation relations of the creation and annihilation operators,
the conditions governing the Bogoliubov coefficients can be
obtained as
\begin{equation}\label{BogoliuCondition}
        \sum_{i_2}\!\left[ \alpha_{i_2i_1}^*\alpha_{i_2j_1}\!-\beta_{i_2i_1}\beta_{i_2j_1}^*\right]
        \!=\!\delta_{i_1j_1},
\quad
        \sum_{i_1}\!\left[ \alpha_{i_2i_1}\alpha_{j_2i_1}^*\!-\beta_{i_2i_1}\beta_{j_2i_1}^*\right]
        \!=\!\delta_{i_2j_2},
\quad
        \sum_{i_1}\! \alpha_{j_2i_1}\beta_{i_2i_1}\!=\!\sum_{i_1}\!\alpha_{i_2i_1}\beta_{j_2i_1},
\quad
        \sum_{i_2}\! \alpha_{i_2j_1}\beta_{i_2i_1}^*\! =\!\sum_{i_2}\!\alpha_{i_2i_1}\beta_{i_2j_1}^*.
\end{equation}
Thereupon, using the vacuum definition
$\hat{a}_{i_2}^{(2)}|0^{(2)}>=0 $, and employing the expansion of
the vacuum state in frame $(2)$ in terms of the Fock space of the
vacuum state in frame $(1)$, we have
\begin{equation}\label{VacuumKExpan}
    \hat{a}_{i_2}^{(2)}|0^{(2)}>=\hat{a}_{i_2}^{(2)}\sum_{n=0}^{\infty}\frac{\hat{a}_{i_{1,1}}^{(1)\dagger}
        \cdots\hat{a}_{i_{1,n}}^{(1)\dagger}}{n!}|0^{(1)}>
 <n^{(1)}|0^{(2)}>=0,
\end{equation}
where $<n^{(1)}|\equiv
<0^{(1)}|\hat{a}_{i_{1,1}}^{(1)}\cdots\hat{a}_{i_{1,n}}^{(1)} $.
Then, utilizing \cref{VacuumKExpan,FrameiToFrameI}, we obtain the
isomorphism $U$ as~\cite{DeWitt-2014}
\begin{equation}\label{VacuumDefined}
    U \: : \: |0^{(1)}>\rightarrow|0^{(2)}>=  U|0^{(1)}>=  C_{2\arrowvert1}\hat{\Psi}^{2\arrowvert1}|0^{(1)}>,
\end{equation}
where
\begin{equation}\label{PsiDefin}
    \hat{\Psi}^{2\arrowvert1}\equiv\exp\left[ \frac{1}{2}\sum\limits_{i_2,i_1,j_1}\beta_{i_2j_1}^*\alpha_{i_2i_1}^{*-1}
    \hat{a}_{i_1}^{(1)\dagger}
    \hat{a}_{j_1}^{(1)\dagger}\right]
\end{equation}
and, due to the first relation of Rels. \eqref{BogoliuCondition},
the inverse of the matrix $\alpha_{i_2i_1}^{*} $ exists and is
defined in such a way that
\begin{equation}\label{InversCoeffi}
    \sum_{i_2}\alpha_{i_2i_1}^{*-1}\alpha_{i_2j_1}^{*}=\delta_{i_1j_1}\qquad\quad\text{and}\qquad\quad
    \sum_{i_1}\alpha_{i_2i_1}^{*-1}\alpha_{j_2i_1}^{*}=\delta_{i_2j_2}.
\end{equation}

At this stage, to determine the normalization constant (or, the
vacuum persistence amplitude~\cite{DeWitt-2014}),
$C_{2\arrowvert1} $, we have
\begin{equation}\label{Normalization}
        <0^{(2)}|0^{(2)}>=
        |C_{2\arrowvert1}|^2<0^{(1)}|\sum_{n=0}^{\infty}\frac{1}{(n!)^2} \sum\limits_{\substack{i_2,j_2, \\
                i_1,j_1,k_1,l_1}}\left(\frac{\beta_{i_2j_1}\alpha_{i_2i_1}^{-1}\hat{a}_{i_1}^{(1)}
            \hat{a}_{j_1}^{(1)}}{2}\right)^{\!\!n}\left(\frac{\beta_{j_2l_1}^*\alpha_{j_2k_1}^{*-1}\hat{a}_{k_1}^{(1)\dagger}
            \hat{a}_{l_1}^{(1)\dagger}}{2} \right)^{\!n}|0^{(1)}>=1.
\end{equation}
Now, using the matrix identity
\begin{equation}\label{MatrixIdentity}
    \det\left( \frac{1}{1-A}\right)\! =\sum_{n=0}^{\infty}\frac{1}{n!}\left[\sum_{m=1}^{\infty}
    \frac{ \text{tr}\left(A^m\right) }{m} \right]^n \quad {\rm :
        when}\ \ A<1,
\end{equation}
for the matrix
$A_{i_1j_1}=\sum_{i_2,j_2,k_1}\beta_{i_2j_1}^*\alpha_{i_2i_1}^{*-1}\beta_{j_2k_1}\alpha_{j_2i_1}^{-1}$
(which, by the first relation of Rels. \eqref{BogoliuCondition},
fulfills the condition $A<1$), and utilizing again the first
relation of Rels. \eqref{BogoliuCondition} and \cref{TotalNumber},
we obtain
\begin{equation}\label{NormalizationValue}
    |C_{2\arrowvert1}|^2=|\det\left( \alpha_{i_2i_1}\right)|^{-1}=\left(1+N^{(2)}+\cdots \right)^{-\frac{1}{2}},
\end{equation}
where $N^{(2)}=\sum_{i_2}N^{(2)}_{i_2}$. Such a result was also
obtained through another approach in
Refs.~\cite{DeWitt-2014,Blaizot-1985}. According to
\cref{VacuumDefined,PsiDefin}, the vacuum state in frame $(2)$ is
equivalent to a statistical state of the particles of the vacuum
state in frame $(1)$. This result represents a generalization of
the Unruh effect for two frames that are accelerating relative to
each other. In this case, the number of particles $N^{(2)} $ is
infinite, and consequently, the vacuum of any two different frames
is unitarily inequivalent or orthogonal to each
other~\cite{DeWitt-2014}, i.e. we have the perpendicularity
$<0^{(1)}|0^{(2)}>=0$. Accordingly, the expected value of an
arbitrary operator, say $\hat{\textbf{O}} $, in frame $(2)$ can
now be written as
\begin{equation}\label{ExpectedValue}
    <\hat{\textbf{O}}>_{(2)}=\frac{<0^{(1)}|\hat{\Psi}^{\dagger2\arrowvert1}
        \hat{\textbf{O}}\hat{\Psi}^{2\arrowvert1}|0^{(1)}>}{<0^{(1)}|\hat{\Psi}^{\dagger2\arrowvert1}
        \hat{\Psi}^{2\arrowvert1}|0^{(1)}>}
    =\frac{\text{tr}\left(\hat{\Psi}^{2\arrowvert1}\hat{\Psi}^{\dagger2\arrowvert1}\hat{\textbf{O}} \right) }
    {\text{tr}\left(\hat{\Psi}^{2\arrowvert1}\hat{\Psi}^{\dagger2\arrowvert1}
    \right)},
\end{equation}
which implies $\pi^{(2)}=U^\dagger\pi^{(1)}U$. In this way, we can
write the correlation function in coordinates $(2)$ in terms of
the Fock space of coordinates $(1)$. On the other hand, it is
obviously known that the correlation function is a key topic of
study in QFT, which can be used to calculate various observables
such as the S-matrix elements. Therefore, we can obtain QFT in
coordinates $(2)$ in terms of the Fock space in coordinates $(1)$.

\section*{Stationary Rigid Reference Frames}
Henceforth, we consider the background to be the Minkowski space
and our task is to describe QFT at any coordinates in this space
in terms of the Fock space in the Minkowski frame. For this
purpose, we first consider a frame from the point of view of a
rigid body rotating around an axis with a uniform proper angular
velocity, say $ \boldsymbol{\omega}$, that accelerates with a
uniform proper acceleration, say acceleration $ \mathbf{a} $, as a
SRRF~\cite{Voytik-2011}. The relation between the coordinates of
this frame, without loss of generality say
$\textbf{x}=(0,\textbf{r}) $, and the coordinates of the flat
Minkowski frame with the Minkowski metric $\eta_{\mu\nu}$, say
$\textbf{X}=(\text{T},\textbf{R}) $, can be written in general as
\begin{equation}\label{LoxodromicFrame}
    \textbf{X}=e^{\left(
        \mathbf{a}.\mathbf{K}+\boldsymbol{\omega}.\mathbf{J}\right)\,t}\,\textbf{x},
\end{equation}
where $ \mathbf{K} $ and $ \mathbf{J} $ are the Lorentz group
generators. We recall that he Minkowski frame can be considered as
a Cauchy surface of the SRRF that evolves by the energy
$\mathbf{a}.\mathbf{K}+\boldsymbol{\omega}.\mathbf{J}$ during the
the SRRF time. In this case, the corresponding metric of this SRRF
is
\begin{equation}\label{LoxodromicMetric}
    ds^2=\left[ \left(\boldsymbol{\omega}\times\mathbf{r} \right)^2\!\!-\!\left(\mathbf{a}.\mathbf{r} \right)^2
    \right] \!dt^2+2\left(\boldsymbol{\omega}\times\mathbf{r}
    \right)\!.d\mathbf{r}dt+d\mathbf{r}^2.
\end{equation}

Now, by \cref{FrameiToFrameI}, the relations between the solutions
$ f_{I\mathbf{a}\boldsymbol{\omega}} $ and $ f_{\mathbf{k}} $, as
well as between the creation (annihilation) operators
$\hat{a}_{I\mathbf{a}\boldsymbol{\omega}}^\dagger\;(\hat{a}_{I\mathbf{a}\boldsymbol{\omega}})$
and $ \hat{b}_{\mathbf{k}}^{\dagger}\;(\hat{b}_{\mathbf{k}}) $,
which belong to the wave equations of the coordinates with energy
$ I $ and the Minkowski coordinates with the momentum $\mathbf{k}
$ and energy $k_0$, respectively, are
\begin{equation}\label{MinToRigid}
        f_{I\mathbf{a}\boldsymbol{\omega}}=\int d^3\mathbf{k}\left( \alpha^{k_0,\mathbf{k}}_{I\mathbf{a}
            \boldsymbol{\omega}}f_{\mathbf{k}}+\beta^{k_0,\mathbf{k}}_{I\mathbf{a}\boldsymbol{\omega}}f_{\mathbf{k}}^{*}\right)
\qquad\quad {\rm and}\qquad\quad
        \hat{a}_{I\mathbf{a}\boldsymbol{\omega}}=\int d^3\mathbf{k}\left( \alpha^{*\, k_0,\mathbf{k}}_{I\mathbf{a}
            \boldsymbol{\omega}}\,\hat{b}_{\mathbf{k}}-\beta^{*\, k_0,\mathbf{k}}_{I\mathbf{a}
            \boldsymbol{\omega}}\,\hat{b}_{\mathbf{k}}^{\dagger}\right),
\end{equation}
where $\alpha^{k_0,\mathbf{k}}_{I\mathbf{a}\boldsymbol{\omega}}$
and $\beta^{k_0,\mathbf{k}}_{I\mathbf{a}\boldsymbol{\omega}}$ and
their conjugates are the Bogoliubov coefficients between these two
coordinates. Also, by isomorphism \eqref{VacuumDefined}, the
relation between the vacuum of these two coordinates is
\begin{equation}\label{RigidVacuumDefined}
    |0_{\mathbf{a}\boldsymbol{\omega}}>=U_{\mathbf{a}\boldsymbol{\omega}}|0_{\eta}>
    =C_{\mathbf{a}\boldsymbol{\omega}}   \hat{\Upsilon}_{\mathbf{a}
        \boldsymbol{\omega}}|0_{\eta}>,
\end{equation}
where $C_{\mathbf{a}\boldsymbol{\omega}} $ is the normalization
constant and
\begin{equation}\label{PsiRigidDefinToMin}
    \hat{\Upsilon}_{\mathbf{a}\boldsymbol{\omega}}\equiv\exp\!\left[\frac{1}{2}\!\sum_{I}\!\int\!\! d^3\mathbf{k}\;
    d^3\mathbf{k}' \beta^{*\, k_0,\mathbf{k}}_{I\mathbf{a}\boldsymbol{\omega}}
    \left(\! \alpha^{*\, k'_0,\mathbf{k}'}_{I\mathbf{a}\boldsymbol{\omega}}\!\right)^{\!-1}
    \hat{b}_{\mathbf{k}}^{\dagger}\,\hat{b}_{\mathbf{k}'}^{\dagger}
    \!\right]
\end{equation}
that consists of a statistical distribution function of the second
quantization operators.

On the other hand, due to \cref{FrameiToFrameI,MinToRigid}, the
Bogoliubov coefficients between any two SRRFs that have different
uniform proper accelerations and different uniform proper angular
velocities are
\begin{equation}\label{BogoCeofRigidRigid}
        \alpha_{I\mathbf{a}\boldsymbol{\omega}}^{I'\mathbf{a}'\boldsymbol{\omega}'}\!\!=\!\!\int\!\!
        d^3\mathbf{k}\left( \alpha^{k_0,\mathbf{k}}_{I\mathbf{a}\boldsymbol{\omega}}\,\alpha^{*\, k_0,
            \mathbf{k}}_{I'\mathbf{a}'\boldsymbol{\omega}'}-\beta^{k_0,\mathbf{k}}_{I\mathbf{a}
            \boldsymbol{\omega}}\,\beta^{*\, k_0,\mathbf{k}}_{I'\mathbf{a}'\boldsymbol{\omega}'}\right)
\qquad {\rm and}\qquad
        \beta_{I\mathbf{a}\boldsymbol{\omega}}^{I'\mathbf{a}'\boldsymbol{\omega}'}\!\!=\!\!\int\!\!
        d^3\mathbf{k}\left( \beta^{k_0,\mathbf{k}}_{I\mathbf{a}\boldsymbol{\omega}}\,\alpha^{k_0,
            \mathbf{k}}_{I'\mathbf{a}'\boldsymbol{\omega}'}-\alpha^{k_0,\mathbf{k}}_{I\mathbf{a}
            \boldsymbol{\omega}}\,\beta^{k_0,\mathbf{k}}_{I'\mathbf{a}'\boldsymbol{\omega}'}\right).
\end{equation}
Obviously, due to the perpendicularly relation, the vacuum of any
two SRRFs that have different uniform proper accelerations and
different uniform proper angular velocities are also orthogonal to
each other. Indeed, utilizing
\cref{NormalizationValue,RigidVacuumDefined}, it can be shown that
when
$(\mathbf{a},\boldsymbol{\omega})\neq(\mathbf{a}',\boldsymbol{\omega}')
$, we have
\begin{equation}\label{RigidNormalConst}
    C^*_{\mathbf{a}\boldsymbol{\omega}}C_{\mathbf{a}'\boldsymbol{\omega}'}\!\!<\!0_{\eta}|
    \hat{\Upsilon}_{\mathbf{a}\boldsymbol{\omega}}^\dagger\hat{\Upsilon}_{\mathbf{a}'
        \boldsymbol{\omega}'}|0_{\eta}\!>=\det \left(\!\alpha_{I\mathbf{a}\boldsymbol{\omega}}^{I'\mathbf{a}'
        \boldsymbol{\omega}'} \right)^{-\frac{1}{2}},
\end{equation}
which, similar to the perpendicularly relation, vanishes.
Therefore, not~only are the vacuum of the Minkowski frame and the
SRRF vacua that have different uniform proper accelerations and
different uniform proper angular velocities perpendicular to each
other, but each of the latter, which can be obtained via the
defined operator
$C_{\mathbf{a}\boldsymbol{\omega}}\hat{\Upsilon}_{\mathbf{a}\boldsymbol{\omega}}$
acting on the Minkowski vacuum, are also orthogonal to each other.
Thus, these vacua form a set of orthogonal vectors for their
corresponding Fock space, which consists of the
$C^*_{\mathbf{a}\boldsymbol{\omega}}\hat{\Upsilon}_{\mathbf{a}\boldsymbol{\omega}}^\dagger$
and
$C_{\mathbf{a}\boldsymbol{\omega}}\hat{\Upsilon}_{\mathbf{a}\boldsymbol{\omega}}$
operators as the creation and annihilation operators. Due to
\cref{PsiRigidDefinToMin,RigidNormalConst}, these operators have
the relation
\begin{equation}\label{RigidCommutator}
    C^*_{\mathbf{a}\boldsymbol{\omega}}C_{\mathbf{a}'\boldsymbol{\omega}'}<0_{\eta}|
    \left[\hat{\Upsilon}_{\mathbf{a}\boldsymbol{\omega}}^\dagger,\hat{\Upsilon}_{\mathbf{a}'
        \boldsymbol{\omega}'}\right]|0_{\eta}>=\delta_{\mathbf{a}\boldsymbol{\omega},\mathbf{a}'\boldsymbol{\omega}'}.
\end{equation}
Consequently, through this approach, we have established
$U_{\mathbf{a}\boldsymbol{\omega}}$-operators that act like a QAO
and transform one vacuum into another. From this point of view,
the $U_{\mathbf{a}\boldsymbol{\omega}}$ operator is a QAO.

\section*{Arbitrary Diffeomorphism of Minkowski Vacuum}
Up to here, we have discussed QFT in a particular frame, the SRRF.
To generalize it to a general frame, we consider a general
diffeomorphism of the Minkowski coordinates. Similar to the
previous section, by \cref{FrameiToFrameI}, the relations between
the solutions of the wave equation of such general coordinates
with energy $I$ and the Minkowski coordinates with the momentum
$\mathbf{k} $ and energy $k_0$ are
\begin{equation}\label{MinToFrameI}
        f_I=\int d^3\mathbf{k}\left( \alpha^{k_0,\mathbf{k}}_{I}f_{\mathbf{k}}+\beta^{k_0,
            \mathbf{k}}_{I}f_{\mathbf{k}}^{*}\right)
\qquad\quad {\rm and}\qquad\quad
        \hat{a}_I=\int d^3\mathbf{k}\left( \alpha^{*\, k_0,\mathbf{k}}_{I}\,\hat{b}_{\mathbf{k}}
        -\beta^{*\, k_0,\mathbf{k}}_{I}\,\hat{b}_{\mathbf{k}}^{\dagger}\right).
\end{equation}
Also, by isomorphism \eqref{VacuumDefined}, the relation between
the vacuum of these two coordinates is
\begin{equation}\label{ArbitrVacuumDefined}
    |0_\text{G}>= C_\text{G}\hat{\Theta}|0_{\eta}>,
\end{equation}
where $C_\text{G}$ is the normalization constant and
\begin{equation}\label{PsiDefinToMin}
    \hat{\Theta}\equiv\exp\!\left[\! \frac{1}{2}\sum_{I}\!\int \!\! d^3\mathbf{k}\; d^3\mathbf{k}'
    \beta^{*\, k_0,\mathbf{k}}_{I}\!\left( \alpha^{*\, k'_0,\mathbf{k}'}_{I}\right)^{\!-1}
    \hat{b}_{\mathbf{k}}^{\dagger}\,\hat{b}_{\mathbf{k}'}^{\dagger}
    \right]
\end{equation}
that again contains a statistical distribution function of the
second quantization operators.

There are two points concerning the $ \hat{\Theta} $ operator. The
first point is that, as mentioned, the set containing
$C_{\mathbf{a}\boldsymbol{\omega}}\hat{\Upsilon}_{\mathbf{a}\boldsymbol{\omega}}$
operators as QAOs forms an orthogonal vector set. Also, due to the
perpendicularly relation, the vacuum of any two different general
diffeomorphisms are perpendicular to each other. Accordingly, this
means that the operator $ \hat{\Theta} $ cannot be written as a
superposition of the operators
$\hat{\Upsilon}_{\mathbf{a}\boldsymbol{\omega}} $, because
otherwise $ <0_{\mathbf{a}\boldsymbol{\omega}}|0_\text{G}>\neq0 $.

The second point is that the operator $ \hat{\Theta} $  can be
written as a pure state of the operators
$\hat{\Upsilon}_{\mathbf{a}\boldsymbol{\omega}} $. To prove this
statement, some explanations are necessary. Using Rels.
\eqref{BogoliuCondition}, it is clear that the function
\begin{equation}\label{SymmetFuncti}
    Q(\mathbf{k},\mathbf{k}')\equiv \sum_{I} \beta^{*\, k_0,\mathbf{k}}_{I}\left( \alpha^{*\, k'_0,\mathbf{k}'}_{I}\right)^{\!-1}
\end{equation}
(as a part of the power of \cref{PsiDefinToMin}) is a symmetric
function with respect to $ \mathbf{k} $ and $\mathbf{k}'$. In the
same way, the function
\begin{equation}\label{RindSymmetrFunct}
    K(\mathbf{k},\mathbf{k}';\mathbf{a},\boldsymbol{\omega})\equiv
    \sum_{I} \beta^{*\, k_0,\mathbf{k}}_{I\mathbf{a}\boldsymbol{\omega}}\left( \alpha^{*\, k'_0,\mathbf{k}'}_{I\mathbf{a}
        \boldsymbol{\omega}}\right)^{\!-1}
\end{equation}
(as a part of the power of \cref{PsiRigidDefinToMin}) is also a
symmetric function with respect to $ \mathbf{k} $ and
$\mathbf{k}'$. Now, due to the first kind Fredholm integral
equation~\cite{Fredholm-1903}, we can write
\begin{equation}\label{Fredholm}
    Q(\mathbf{k},\mathbf{k}')=\int
    d^3\mathbf{a}\,d^3\boldsymbol{\omega}\, h(\mathbf{a},\boldsymbol{\omega})\,
    K(\mathbf{k},\mathbf{k}';\mathbf{a},\boldsymbol{\omega}),
\end{equation}
with $K(\mathbf{k},\mathbf{k}';\mathbf{a},\boldsymbol{\omega})$ as
a kernel function and $h(\mathbf{a},\boldsymbol{\omega})$ as a
weight function. Accordingly, we can obtain
\begin{equation}\label{ArbitraryRindler}
    \hat{\Theta}=
    \prod_{\mathbf{a},\boldsymbol{\omega}\in\, \Omega} \left( \hat{\Upsilon}_{\mathbf{a}\boldsymbol{\omega}}\right)^{\epsilon},
\end{equation}
where $ \epsilon\equiv d^3\mathbf{a}\,d^3\boldsymbol{\omega} $ and
$\Omega $ is the domain of uniform proper acceleration and uniform
proper angular velocity, which contains the weight function
$h(\mathbf{a},\boldsymbol{\omega}) $. \cref{ArbitraryRindler}
indicates that the operator $\hat{\Theta} $ forms a QAO.
Consequently, in this way, the
$C_{\mathbf{a}\boldsymbol{\omega}}\hat{\Upsilon}_{\mathbf{a}\boldsymbol{\omega}}$
operators form an orthonormal basis such that the vacuum of any
coordinates can be considered as a pure state of these operators.

At last, under these explanations, we have been able to add a set
of vectors
$C_{\mathbf{a}\boldsymbol{\omega}}\hat{\Upsilon}_{\mathbf{a}\boldsymbol{\omega}}|0_{\eta}>$
to the Hilbert space to obtain an augmented Hilbert space. Now,
utilizing such an augmented Hilbert space, QFT can be described in
a general frame and some kind of quantum expression for
acceleration can be researched.

\section*{Metric Field Extraction}
Based on previous sections, we have shown how the expectation
value of an operator (in particular, the correlation function) can
be described based on the augmented Hilbert space. Now, in this
section, we use this result to extract the metric field of a general frame of the
Minkowski spacetime.

At first, according to Refs.~\cite{Saravani-2016,Perche-2022},
from a two-point correlation function, say $W\left(x,x' \right)$,
a metric field can be extracted as
\begin{equation}\label{MetricDefine}
    g_{\mu\nu}(x)=\lim\limits_{x'\rightarrow x}\left\lbrace -\frac{1}{8\pi^2}\partial_\mu\partial'_{\nu}
    \left[ W\left(x,x' \right)\right]^{-1}  \right\rbrace.
\end{equation}
In Refs.~\cite{Saravani-2016,Perche-2022}, the approach employed
to determine a two-point correlation function is to use the wave
equation through a specific metric in obtaining its solution
similar to \cref{GeneralField}, namely
\begin{equation}\label{Two-PointCorrelation}
    W\left(x,x' \right)=<0_\text{G}|\hat{\Phi}(x)\hat{\Phi}(x')|0_\text{G}>.
\end{equation}
Thereafter, by inserting the obtained solution into this relation,
they achieve the two-point correlation function
\eqref{Two-PointCorrelation} as
\begin{equation}\label{Two-PoiCorr}
    W\left(x,x' \right)=\sum_{I,J}<0_\text{G}|\left[
    \hat{a}_{I}f_{I}(x)+\hat{a}^{\dagger}_{I}f^*_{I}(x)\right]
    \left[
    \hat{a}_{J}f_{J}(x')+\hat{a}^{\dagger}_{J}f^*_{J}(x')\right]|0_\text{G}>
    =\sum_{I}f_{I}(x)f^*_{I}(x').
\end{equation}
Then, by substituting this relation into \cref{MetricDefine}, they
obtain the metric. Such an approach is also used to define local
quantum clocks and rulers. Accordingly, the attempt to explain
nature is to use two fundamental concepts, i.e. the classical
metric field and the Hilbert space in quantum mechanics, which
are~not compatible with each other.

Instead, in the approach presented in this work, we replace the
traditional fundamental concepts of physics with another
fundamental concept, the augmented Hilbert space. In this way,
utilizing the obtained operator \eqref{ArbitraryRindler} and
vacuum \eqref{ArbitrVacuumDefined}, we obtain the two-point
correlation function as
\begin{equation}\label{Two-PoiCorrValue}
    W\left(x,x' \right)=\int d^3\mathbf{k}d^3\mathbf{k}'<0_\text{G}|\left(
    \hat{b}_{\mathbf{k}}f_{\mathbf{k}}+\hat{b}^{\dagger}_{\mathbf{k}}f^*_{\mathbf{k}}\right)
    \left(
    \hat{b}_{\mathbf{k}'}f_{\mathbf{k}'}+\hat{b}^{\dagger}_{\mathbf{k}'}f^*_{\mathbf{k}'}\right)|0_\text{G}>.
\end{equation}
Then using \cref{MinToFrameI,ArbitrVacuumDefined,PsiDefinToMin},
we can get
\begin{equation}\label{DetailProof}
    \begin{split}
        <0_\text{G}|
        \hat{b}_{\mathbf{k}}
        \hat{b}_{\mathbf{k}'}|0_\text{G}>=&\sum_{I}\alpha^{k_0,\mathbf{k}}_{I} \beta^{*\,
        k'_0,\mathbf{k}'}_{I},
        \qquad\qquad
        <0_\text{G}|
        \hat{b}_{\mathbf{k}}
        \hat{b}_{\mathbf{k}'}^\dagger|0_\text{G}>=&\sum_{I}\alpha^{k_0,\mathbf{k}}_{I} \alpha^{*\,
        k'_0,\mathbf{k}'}_{I},
        \\
        <0_\text{G}|
        \hat{b}_{\mathbf{k}}^\dagger
        \hat{b}_{\mathbf{k}'}|0_\text{G}>=&\sum_{I}\beta^{k_0,\mathbf{k}}_{I} \beta^{*\,
        k'_0,\mathbf{k}'}_{I},
        \qquad\qquad
        <0_\text{G}|
        \hat{b}_{\mathbf{k}}^\dagger
        \hat{b}_{\mathbf{k}'}^\dagger|0_\text{G}>=&\sum_{I}\beta^{k_0,\mathbf{k}}_{I}
        \alpha^{*\, k'_0,\mathbf{k}'}_{I}.
    \end{split}
\end{equation}
Now, by inserting these relations into \cref{Two-PoiCorrValue},
while utilizing the inversion of \cref{MinToFrameI}, we reach the
final result of \cref{Two-PoiCorr} for the two-point correlation
function. Then, by substituting this final result into
\cref{MetricDefine}, we obtain the desired metric, a quantum
expression as an alternative to the classical concept of the
metric field. Consequently, by using \cref{MetricDefine} obtained
from this approach, it is possible to extract features of
spacetime from the augmented Hilbert space.

Therefore, we are able to extract the metric of a general
diffeomorphism of the Minkowski coordinates using only the tools
provided in the standard QFT in the Minkowski coordinates, i.e.
the second quantization operators $ \hat{b}_{\mathbf{k}} $ and
$\hat{b}^{\dagger}_{\mathbf{k}} $, the operators
$\hat{\Upsilon}_{\mathbf{a}\boldsymbol{\omega}} $ and
$\hat{\Upsilon}_{\mathbf{a}\boldsymbol{\omega}}^\dagger $ as QAOs
(as statistical distribution functions of the second quantization
operators), and the Minkowski vacuum $ |0_\eta> $. Accordingly,
the Minkowski spacetime in arbitrary frame and its metric field in
this frame only emerge from the vectors of the augmented Hilbert
space. In fact, although we started from the classical metric, we
can now replace it with a metric that can be extracted from the
obtained operators and the augmented Hilbert space. In this way,
through the augmented Hilbert space, we remedy the
incompatibilities between the classical metric field and the
Hilbert space in quantum mechanics.

The following chart briefly summarizes what we have performed in
the presented approach.
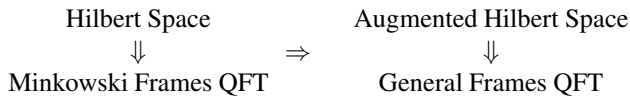
\begin{equation*}\label{Chart}
    \begin{array}{cc}
        & \text{Hilbert Space} \\
        & \Downarrow \\
        & \text{Minkowski Frames QFT}
    \end{array}
    \!\Rightarrow\!
    \begin{array}{cc}
        & \text{Augmented Hilbert Space} \\
        & \Downarrow \\
        & \text{General Frames QFT}
    \end{array}
\end{equation*}

\section*{Conclusions}
By explaining some ambiguities and issues of spacetime and metric
field, as fundamental concepts in physics, we have discussed the
need for an alternative fundamental concept. Then, from the Unruh
effect point of view and using the GNS construction, we have
investigated the properties and relations between the vacuum of
two different coordinates. In this regard, one of the important
properties is that the vacuum of two different frames that are
accelerating relative to each other are orthogonal. Hence, one can
write the correlation function, and in turn QFT, in one frame in
terms of the Fock space of another frame.

Then, in the Minkowski space, by choosing the Minkowski frame and
any SRRF, we have determined the relation between the vacuum of
these two coordinates and shown that the vacuum of any SRRF (as an
accelerated frame) can be obtained by applying the defined
operator to the vacuum of the Minkowski frame. The mentioned
operator consists of a statistical distribution function of the
second quantization operators of the Minkowski frame via the
Bogoliubov coefficients between these two coordinates. Also,
utilizing the Bogoliubov coefficients, we have indicated that
not~only are the vacuum of the Minkowski frame and the SRRF vacua
that have different uniform proper accelerations and different
uniform proper angular velocities perpendicular to each other, but
each of the latter are also orthogonal to each other. Thus, these
vacua form a set of orthogonal vectors for their corresponding
Fock space composed of those defined operators. Since such
operators transform one vacuum into another, we have referred to
them as QAOs, which form an orthogonal basis.

Thereupon, we have generalized these results to a general
diffeomorphism of the Minkowski coordinates, and hence the
corresponding operator can be written as a pure state of that
orthogonal basis. Thus, the vacuum of any general frame in the
Minkowski spacetime can also be represented as a pure state of
SRRFs. In this way, we have illustrated that the mechanism of the
Unruh effect leads, mathematically, to a quanta of acceleration;
and, acceleration is~not a state of particles, but an independent
quantity.

By adding the QAOs to the Hilbert space, we have obtained the
augmented Hilbert space that includes QFT in a general frame.
Finally, given that the metric field can be obtained from a
two-point correlation function, we have been able to extract the
metric of a general diffeomorphism of the Minkowski coordinates
using only the tools provided in the augmented Hilbert space (i.e.
the second quantization operators, the QAOs and the Minkowski
vacuum). Consequently, it is possible to extract features of
spacetime also from the augmented Hilbert space. On this basis,
the diffeomorphism of the Minkowski spacetime and its related
metric field only emerge from the augmented Hilbert space vectors.
However, our analysis is limited to diffeomorphic images of the
Minkowski spacetime that has been used as a fiducial metric. In
fact, although we have started from the classical metric, we have
now replaced it with a metric that can be extracted from the
obtained operators and the augmented Hilbert space. In this way,
we have replaced the discussed traditional fundamental concepts of
physics with another fundamental concept, the augmented Hilbert
space.

\section*{Acknowledgements}

We thank the Research Council of Shahid Beheshti University.


%
\end{document}